\documentclass[twocolumn,showpacs,preprintnumbers,amssymb,showkeys]{revtex4}

\usepackage{graphicx}

\begin{document}
\title{The influence of the weak bond-energy dimerization 
on the single-particle 
optical conductivity  of quasi-one-dimensional systems}
\author{Ivan Kup\v{c}i\'{c}} 
  \affiliation{ Department of Physics, 
         Faculty of Science, POB 331,  HR-10 002 Zagreb, Croatia}

\begin{abstract}
The single-particle contributions to the optical conductivity of the
quasi-one-dimensional systems has been reexamined by using
the gauge-invariant transverse microscopic  approach.
	The valence electrons  are described by a model with the 
weak bond-energy dimerization,  while the relaxation processes are
taken into account phenomenologically.
	It turns out that the interband conductivity of the 
insulating half-filled case fits well the single-particle
optical conductivity measured in various CDW systems.
	In the metallic regime, for the doubled Fermi vector
$2k_{\rm F}$ close to $\pi/a$, the conduction electrons exhibit a non-Drude
low-frequency response, with the total spectral weight
shared between the intraband and interband channels nearly in equal proportions.
 	For $2k_{\rm F} - \pi/a$ not to small,  
the behaviour of the conduction electrons  can be described 
as the response of a simple Drude metal.
\end{abstract}


\maketitle

\section{Introduction}
In recent years there has been a lot of experimental
\cite{Jacobsen1,Jacobsen2,Degiorgi1,Degiorgi2,Schwartz,Uchida,Lupi}
and theoretical 
\cite{Kim,Emery,Kupciccm}
interest in low dimensional  systems which
exhibit a gap (for $T$ below the critical temperature $T_c$)
or pseudogap ($T > T_c$) features in the low-frequency optical conductivity.
	In principle, it is possible to solve the single-particle 
contributions to the response
functions for the ideal electronic system and then to take the single-particle
damping phenomenologically.
	However, a good agreement with experiments
cannot be achieved by such phenomenological extension of 
the standard transverse microscopic theory,
which was the question raised by Kim at al.  \cite{Kim}.
	Instead, they developed a more rigorous method
which fits well the single-particle conductivity in various
quasi-one-dimensional (Q1D) systems with the CDW instability.

Surprisingly, a straightforward calculation gives that the longitudinal
microscopic response theory, with the phenomenologically treated disorder, also
agrees with the single-particle optical conductivity measured  in the CDW systems
\cite{Kupcicup}.
	The origin of the inconsistency between two microscopic approaches
represents an interesting question which will be examined in the present 
paper.
	The purpose of the paper is threefold;
	first, to determine the gauge-invariance requirement for the electronic 
systems with the bond-energy dimerization 
(by considering the simplest case);
	second, to argue that in the weak-splitting limit the dielectric function
has the usual form, at variance with the conclusions of the analysis
based on the electric-dipole approximation \cite{ZBB}
that the dielectric function
has the Lorentz-Lorenz form even in this limiting case;
	finally, to illustrate that in the weak-splitting limit various 
Drude-like and non-Drude low-frequency metallic behaviours are possible.

In this paper, we shall consider the electronic system  stabilized
into an ordered phase described by the wave vector $Q = \pi /a$.
	It will be assumed that $Q $
differs from the doubled Fermi vector, $2k_{\rm F}$, 
leading to a metallic state with the bond-energy dimerization.
	Such assumption is natural, in particular
in the strong-splitting case, describing primarily the molecular crystals
with an important role played by the local field corrections.
	According to Refs. \cite{Brazovski1,Brazovski2}, 
the weak-splitting $Q \neq 2k_{\rm F}$ microscopic
model should also be interesting.
	The optical conductivity obtained in this case is expected to be 
relevant to the CDW systems, both at $T < T_c$  and $T > T_c$, but for 
the frequencies well above the frequency of the pinned collective mode.

In Sec. 2 we determine the optical conductivity
$\sigma_{\alpha}^{\rm total } ( \omega)$ 
and the associated dielectric
function in the Q1D model with the weak bond-energy dimerization,
taking properly into account gauge invariance of the response theory.
	Sec. 3 illustrates that the gauge invariance enables the rigorous
treatment of the effective mass theorem (and the associated effective
number of conduction electrons), as well as of the conductivity sum rules.
	Additionally, it is explained how the standard transverse treatment 
of the interband optical processes \cite{LRA,Gruner}
can be remedied.
	The development of 
$\sigma_{x}^{\rm total } ( \omega)$ 
with doping is shown for a few characteristic cases.
	Finally, the concluding remarks are given in Sec. 4.

\section{Theoretical model}

We consider the Q1D 
electronic system with the bond-energy 
dimerization along the $x$ axis
in the presence of the electromagnetic fields.
	The Hamiltonian reads as
\begin{eqnarray}
H &=&  
\sum_{n \sigma} [
t_> \tilde{\beta}^{\dagger}_{n  \sigma} \tilde{\alpha}_{n \sigma}   
+  t_< \tilde{\alpha}^{\dagger}_{n+2a  \sigma} \tilde{\beta}_{n \sigma}  
\label{eq1}\\ \nonumber \
&& + t_b (\tilde{\alpha}^{\dagger}_{n+ b  \sigma}  \tilde{\alpha}_{n   \sigma}
+
\tilde{\beta}^{\dagger}_{n +b \sigma}  \tilde{\beta}_{n   \sigma} ) 
+  {\rm H.c.}],
\end{eqnarray}
where the influence of the external fields is introduced through
\begin{eqnarray}
\tilde{l}^{\dagger}_{n \sigma} &=& l^{\dagger}_{n \sigma}
e^{\mathrm{i}e  /(\hbar c) ({\bf R}_n + {\bf r}_l) \cdot  
{\bf  A} ( {\bf R}_n + {\bf r}_l )}, 
\label{eq2}\end{eqnarray}
and where $l^{\dagger}_{n \sigma}$, $l \in \{ \alpha, \beta \}$,
is a creation operator on the  ${\bf R}_n + {\bf r}_l$ site.
	In order to preserve a simple form of the intracell
phase factors, we approximate
the positions of  atoms within 
the unit  cell of the dimerized case  with their positions in
 the original lattice (for example,
$r_{ \alpha x} = 0$ and $r_{ \beta x} = a$,
with ${\bf a}_1 = a \hat{{\bf x}}$ and ${\bf a}_2 = b \hat{{\bf y}}$
representing the primitive vectors of the original lattice).
	For the bond energies we assume that $t_i < 0$
and $|t_>| > |t_<| > |t_b|$,
i.e., the $x$ axis represents the strong axis of the present Q1D system.

In principle, it is possible to generalize the theory to include all 
relevant relaxation processes explicitly.
	Instead, we adopt an approach which assumes that in the clean 
(i.e. underdamped) limit
one can separate the relaxation of the electron momentum on the static 
dimerization potential (or on the (quasi)static $Q = \pi /a$ phonon modes)
from other electron scattering processes (related to the impurities and 
other phonon modes), calculate the dimerization 
(or (quasi)dimerization) effects exactly, and,
finally, take the remaining relaxation effect into account phenomenologically.

\subsection{Bare Hamiltonian}

	The Bloch operators of the dimerized case 
satisfy the relation
\begin{eqnarray}
L^{\dagger}_{{\bf k} \sigma} &=& \frac{1}{\sqrt{N}} \sum_{n} 
e^{ \mathrm{i}{\bf k} \cdot {\bf R}_n} \sum_{l }
 e^{\mathrm{i}  k_x  r_{lx}}
U_{ k_x} (L, l) l^{\dagger}_{n \sigma}, \nonumber \\
&& 
\label{eq3}\end{eqnarray}
where  $e^{i  k_x r_{lx}}$ are the bare intracell
phase factors, and where  $U_{k_x} (L, l)$ represent the 
transformation-matrix elements dependent on an additional phase
$\varphi (k_x)$, as discussed below.
	For $t_i < 0$, 
the band indices $L=S$ and $L=A$ stand for the lower (symmetric) and
upper (antisymmetric) band, respectively.
	It is also assumed that 
the lower band  is partially filled and the upper band is empty.
	If the Fermi level is in the upper band, the following expressions
remain the same, only the electron picture has to be replaced by the 
hole picture.

Inserting ${\bf A} ({\bf R}_n + {\bf r}_l) = 0$ in Eqs. (\ref{eq1}) 
and (\ref{eq2}), 
we obtain the bare Hamiltonian $H_0$.
	After diagonalization, this Hamiltonian becomes
\begin{eqnarray}
H_0 &=& \sum_{{\bf k} \sigma L} E_L ({\bf k}) 
L^{\dagger}_{{\bf k} \sigma} L_{{\bf k} \sigma}. 
\label{eq4}\end{eqnarray}
	The Bloch energies are given by
\begin{eqnarray}
E_{A,S} ({\bf k}) &=& E_b ({ k_y})\pm E_a ({k_x}), 
\label{eq5}\end{eqnarray}	
where
\begin{eqnarray}
E_b ( k_y) &=& -2 |t_b| \cos k_y b ,
\nonumber \\
E_a ( k_x) &=&  \sqrt{\varepsilon^2 ( k_x) + \Delta^2 (k_x)}.
\label{eq6}\end{eqnarray}	
	The corresponding transformation-matrix elements are
\begin{eqnarray} 
&& \left( \begin{array}{ll} 
  U_{k_x} (S,\alpha) & U_{k_x} (S,\beta)  \\
  U_{k_x} (A,\alpha) & U_{k_x} (A,\beta) 
\end{array} \right)   
=  \nonumber \\
&& 
= \frac{\sqrt{2}}{2}\left( \begin{array}{cc}
  e^{\mathrm{i}\varphi ({k_x})/2 } &  e^{-\mathrm{i}\varphi ({k_x})/2}   \\
  e^{\mathrm{i}\varphi ({k_x})/2} &  -e^{-\mathrm{i}\varphi ({k_x})/2}    
\end{array} \right),  
\label{eq7}\end{eqnarray} 
where the auxiliary phase $\varphi (k_x)$ 
satisfies the $p$-symmetry-gap relation
\begin{eqnarray}
\tan \varphi (k_x) &=&  \frac{\Delta ( k_x)}{\varepsilon ({k_x})}.
\label{eq8}\end{eqnarray} 
	Other notation is given by
\begin{eqnarray}
\varepsilon ({\bf k}) &=& 
2|t_a| \cos  k_x   a,
\nonumber \\
\Delta ({\bf k}) &=& \Delta_0  \sin  k_x a,
\label{eq9}\end{eqnarray}
with $2|t_a| \equiv |t_>| + |t_<|$ being the averaged intrachain bond
energy and $\Delta_0 \equiv |t_>| - |t_<|$ 
is the magnitude of the gap function.

One recognizes in Eqs. (\ref{eq3}) and (\ref{eq7}) 
the total intracell phases of the form
$\Phi_{\alpha} (k_x) = k_x r_{\alpha} + \varphi (k_x)/2$, for 
the $| \alpha n \rangle$ states, and
$\Phi_{\beta} (k_x) = k_x r_{\beta} - \varphi (k_x)/2$,
for the $| \beta n \rangle$ states.
	At this point it is interesting to note that if we take the 
strong-splitting limit, where $\Delta_0/(2|t_a|) \approx 1$,
we obtain $\varphi ( k_x) \approx k_x a$, and finally
$\Phi_{\alpha} (k_x) \approx \Phi_{\beta} (k_x)$.
	Such structure of the intracell phases suggests the usage
of the molecular representation, based on 
the molecular orbitals 
$1/\sqrt{2}(| \alpha n \rangle \pm | \beta n \rangle)$,
with a further electric-dipole approximation (for example, see
Ref. \cite{ZBB}).
	In the weak-splitting limit, however,
the atomic $| l n \rangle$ representation is more adequate, as argued
in the below discussion of the interband current vertices.

\subsection{Electron-photon coupling}
In order to write explicitly all relevant electron-photon
coupling functions, we need the Taylor expansion in the vector potential  
of Eq. (\ref{eq1}) to the second order.
	The first-order  term and the direct 
second-order term become
\begin{eqnarray}
H^{{\rm ext}}_{1} &=& -\frac{1}{c}  
\sum_{{\bf k} \sigma} \sum_{LL'} \sum_i [A_{x_i} ( {\bf q}_{\perp i})
J^{LL'}_{x_i} ( {\bf k}) 
\nonumber \\ 
&&\times
L^{\dagger}_{{\bf k} + {\bf q}_{\perp i}  \sigma} L'_{{\bf k} \sigma}  
 + {\rm H.c.}],
\nonumber \\
H^{{\rm ext}}_{2} &=& \frac{e^2}{2mc^2}  
\sum_{{\bf k} \sigma} \sum_{LL'} \sum_i [A^2_{x_i} ( {\bf q}_{\perp i})
\gamma^{LL'}_{x_i x_i} ( {\bf k};2) 
\nonumber \\ 
&\times&
L^{\dagger}_{{\bf k} + {\bf q}_{\perp i}  \sigma} 
L'_{{\bf k} \sigma} 
+ {\rm H.c.}], 
\label{eq10}
\end{eqnarray}
respectively,
with $x_i \in \{ x, y \}$.
	Furthermore, $A_{x_i} ( {\bf q}_{\perp i})$ 
and $A^2_{x_i} ( {\bf q}_{\perp i})$ 
represent the Fourier transforms of $A_{x_i} ( {\bf r})$ 
and $A^2_{x_i} ( {\bf r})$, and \hbox{${\bf q}_{\perp i} \cdot {\bf a}_i = 0$.}
	The explicit form of the coupling functions, 
the current vertices $J^{L_1L_2}_{x_i} ( k_i)$
and the bare Raman vertices $\gamma ^{^{L_1L_2}}_{x_ix_i} ( k_i;2)$,
is given in the Appendix.

It is essential to notice that in the bond-energy dimerization model
two physically different terms comprise
the interband current vertices.
	The first term is dispersionless and corresponds 
to the intramolecular
interband processes (between the states $| \alpha n \rangle$
and $| \beta n \rangle$).
	The second term describes all other, intermolecular
interband processes.
	In the strong-splitting limit, one obtains the 
interband current vertices consisting mainly of the intramolecular contribution,
$J_x^{SA} (k_x) \approx \mathrm{i} e a |t_>| /\hbar$, 
giving the associated dipole vertex of the form
$  e a/2$.
	Although in this case the response functions are complicated functions
of the short-range dipole-dipole interactions, it is possible to apply
the electric-dipole approximation, and to obtain
the Q1D optical model extensively studied 
 by \v{Z}upanovi\'{c} at al. \cite{ZBB}.
	In the present article we want to extract physical 
information from the optical conductivity determined 
in the opposite (weak-splitting) limit.
	In this case the intramolecular and intermolecular interband processes
are equally important, and, consequently, the local field corrections become 
negligible.

\subsection{Optical conductivity in the weak-splitting limit}
To obtain  the gauge invariant form
of the (transverse) optical conductivity in the ideal multiband models,
it is necessary
first to gather the diamagnetic photon self-energy with
the real part of the static interband photon self-energies
\cite{Pines,Mahan}.
	This results in the effective mass theorem
\cite{Kupciccm} 
\begin{eqnarray} 
[m^{SS}_{x_ix_i} ( k_{x_i})]^{-1} & =& 
\frac{1}{m} \gamma^{SS}_{x_ix_i} ( k_{x_i}) 
= \frac{1}{m} \gamma^{SS}_{x_ix_i} ( k_{x_i};2)
\nonumber \\
& &- 
\frac{2|J^{SA}_{x_i} ( k_{x_i})|^2
}{  e^2E_{AS} ( k_{x})}.
\label{eq11}\end{eqnarray}
	Here $E_{AS} ( k_{x}) = E_A ({\bf k}) - E_S ({\bf k}) = 2 E_a (k_{x})$.
	$\gamma^{SS}_{x_ix_i} ( k_{x_i}) $ is the static Raman vertex,
which can also be written in the form 
$(m/\hbar^2) \partial ^2 E_S ({\bf k}) / \partial k_{x_i}^2$
\cite{Abrikosov}.
	When the local field corrections are absent for the symmetry reasons, 
or are negligible,
the remaining  contributions to the total photon self-energy lead
to the ideal intraband and interband optical conductivities 
given by
\begin{eqnarray}
 \sigma_{x_i}^{\rm intra} (\omega, \eta_1) &=& \frac{\mathrm{i}}{\omega}
\frac{e^2 n_{x_i}^{\rm eff}}{m} 
  \frac{ \hbar \omega}{\hbar \omega + \mathrm{i}\eta_1}
  ,  \label{eq12} \\ \nonumber
\sigma_{x_i}^{\rm inter} (\omega, \eta_2) &=& \frac{\mathrm{i}}{\omega}
\frac{1}{V} \sum_{{\bf k} \sigma} \frac{(\hbar \omega)^2
|J_{x_i}^{SA} ( k_{x_i})|^2  
}{E_{AS} ^2 ( k_{x_i}) }
\\ \nonumber
 &&\times 
\bigg\{  \frac{  f_S({\bf k}) 
}{\hbar  \omega  -  E_{AS} ( k_{x_i})    + 
 i \eta_2 } 
\\ &&
 -
\frac{  f_S({\bf k}) 
}{\hbar  \omega + E_{AS} ( k_{x_i})  + 
 \mathrm{i} \eta_2 } 
\bigg\},
\label{eq13}\end{eqnarray}
respectively.
	The effective number of conduction 
electrons can be shown as an integral of the static Raman vertex
over all occupied states
\cite{Jacobsen2,Kupciclong},
 \begin{eqnarray}
n^{\rm eff}_{x_i} &=& \frac{1}{V} \sum_{{\bf k} \sigma} 
\gamma^{SS}_{x_ix_i} ( {\bf k})   f_S({\bf k}) ,
\label{eq14}\end{eqnarray}
or as an integral of the square of the electron group velocity over the Fermi
surface,
\begin{flushleft}
$$
n_{ x_i}^{\rm eff} = m\frac{1}{V} \sum_{{\bf k} \sigma} 
|v_{x_i} ( k_{x_i})|^2 
\delta [E_S ({\bf k}) - \mu], 
\eqno{(14')}
$$
\end{flushleft}
(notice that for the free electrons both of these expressions reduce to the
total number of conduction electrons).
	Here $ f_S({\bf k})$ is the abbreviation for the Fermi-Dirac function
$ f [E_S({\bf k})]$.
	Furthermore, notice that the expressions 
(\ref{eq12}) and (\ref{eq13}) 
describe the case of a normal photon incidence, 
${\bf A} ({\bf r}) = (A_x ( y), 0)$, or the case of the parallel photon
incidence, 
${\bf A} ({\bf r}) = (0, A_y (x))$.
	The generalization of these expressions is straightforward, and 
will be briefly discussed below.

It is rather a good approximation to treat the single-particle damping 
(connected with the impurities and the remaining phonons)
phenomenologically, by replacing the adiabatic terms $\eta_j$
with the corresponding damping energies $\Sigma_{jx_i}$. 
	Then the total conductivity becomes
\begin{eqnarray}
 \sigma_{x_i}^{\rm total} (\omega) & =& 
 \sigma_{x_i}^{\rm intra} (\omega, \Sigma_{1x_i})
\nonumber \\
&& + 
 \sigma_{x_i}^{\rm inter} (\omega, \Sigma_{2x_i}).
\label{eq15}\end{eqnarray} 
	Finally, we obtain that the associated dielectric function 
reads as
\begin{eqnarray}
\varepsilon_{x_i} (\omega) &\approx& \varepsilon_{\infty}
+ \frac{4 \pi \mathrm{i}}{\omega }   \sigma^{\rm total}_{x_i} (\omega).
\label{eq16}\end{eqnarray}
	The contributions of the high-frequency optical processes,
not included in Eq. (\ref{eq1}), are assumed to be constant up to frequencies
well above the considered frequency region,
and are represented by $\varepsilon_{\infty} -1$.

\section{Results and discussion}
In the weak-splitting conductivity (\ref{eq15}),  
$2|t_a|$ is the largest  parameter of the model.
	In spite of the phenomenological description of the single-particle 
damping, this expression is a rather complicated function of 
the remaining parameters,
$2|t_b|$, $k_{\rm B}T$, $\Sigma_{i}$ and  $E_{AS} ({\bf k}_{\rm F})$
(the dependence of $\sigma_{x_i}^{\rm total} (\omega)$
on the parameter $2{\bf k}_{\rm F} - {\bf Q}$ is represented here
through the bare interband threshold energy 
$E_{AS} ({\bf k}_{\rm F})$) 
which makes further rigorous analytic treatment very difficult.
	In this respect, we shall limit the below discussion 
only on some simple limiting cases.

\subsection{Strictly 1D limit}
If the electronic system is treated as strictly one-dimensional, i.e. 
$|t_b| = 0$, one obtains that  the electron dispersion 
in the perpendicular direction, $E_b (k_y)$,  vanishes.
	Similarly, for the incidence of the electromagnetic fields 
normal to the chains, the probability of the photon absorption (or emission)
disappears, as well.
	In this case, the real part of  the conductivity comprises only the 
$\delta ( \omega )$ term,
corresponding to the elastic photon scattering.
	According to the effective mass theorem (\ref{eq11}),
the coupling constant arises from two mutually competing terms,
the diamagnetic term and
the static contributions in the interband term, and is 
given by the $\eta_1 \rightarrow 0$ limit of (\ref{eq12}),
with $n^{\rm eff}_{x_i}$ given by (\ref{eq14}).
	For non-normal incidence, the effective number of 
conduction electrons, $n^{\rm eff}_x$, has to be replaced
by $\sin^2 \varphi \; n^{\rm eff}_x$, where $\varphi$ is the angle between
${\bf A} ({\bf r})$ and $\hat{ {\bf x}}$.
	In this way, one can derive the well-known anisotropic dispersion 
of the longitudinal plasma modes.
	Namely, 
\begin{eqnarray}
 \sigma_{x}^{\rm intra} (\omega) & \approx& \frac{\mathrm{i}}{\omega}
\frac{e^2 n_{x}^{\rm eff}}{m}  \sin^2 \varphi,
\label{eq17}\end{eqnarray} 	 
and finally
\begin{eqnarray}
\omega^2_{\rm pl} ({\bf q}) &\approx &  
\frac{4 \pi e^2 n_{x}^{\rm eff}}{\varepsilon_{\infty}m} 
\frac{q_x^2}{q_x^2 + q_y^2}.
\label{eq18}\end{eqnarray}

To proceed with the numerical calculation, we will 
take the conductivity (\ref{eq15}) in the usual Q1D limit,
$|t_a| \gg |t_b| \neq 0$ (hereafter this limit is denoted by 
$|t_b| \rightarrow 0$).
	In the optical conductivity considerations concerned only with the
single-particle contributions, the ``explicit'' influence  of the
interchain hopping processes on Eq. (\ref{eq15}), through 
the Bloch energies, can be neglected. 
	Nevertheless,  these processes 
will be taken into account implicitly, since they 	
lead to a finite
probability of the photon absorption (emission).
	In such circumstances the 1D version of Eq. 
(\ref{eq15}), with the non-zero
damping energies, can  safely be used.
	Furthermore, we restrict the analysis on the 
case of normal photon incidence, where only 	
the $x_i = x$ conductivity is non-zero.
	We also assume that $\Sigma_{1} = \Sigma_2 \equiv \Sigma$
and ${\bf k}_{\rm F} = k_{\rm F} \hat{{\bf x}}$.

\subsection{Gauge invariance}

The gauge invariance of the conductivity (\ref{eq15}) can be easily proved 
term by term, by the
direct calculation of the longitudinal dielectric function,
following the procedure developed in Refs. \cite{Kupciclong} and 
\cite{Kupciccm}.
	The most important qualitative consequences of 
the gauge-invariance requirement are as follows.

    \begin{figure}[tb]
     \includegraphics[height=15pc,width=15pc]{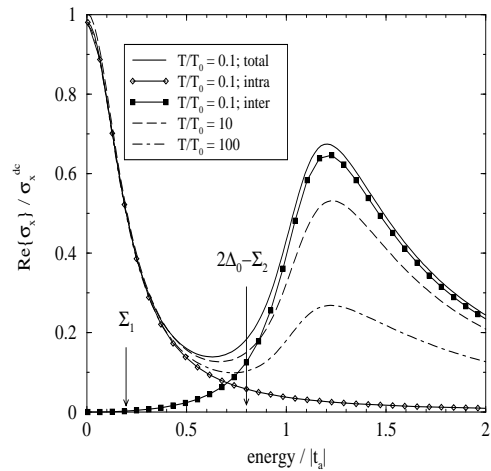}
    \caption{The explicit temperature dependence
    of the real part of the optical conductivity for 
    $|t_b| \rightarrow 0$, $\Delta_0 = 0.5 |t_a|$, 
    $\Sigma = 0.2|t_a|$, and
    $k_{\rm F} = 0.475\pi/a$.
	The temperature scale $T_0$ is given by 
    $k_{\rm B}T_0  = 10^{-3}  \cdot 2|t_a|$ 
    (for example, $T_0 \approx 4$ K,
    $\Delta_0 \approx 0.1$ eV and $\Sigma \approx 0.04$ eV,
    for $|t_a| \approx 0.2$ eV).
    The intraband and interband contributions at $T = 0.1T_0$ are also shown.
    }
     \end{figure}

    \begin{figure}[tb]
     \includegraphics[height=15pc,width=15pc]{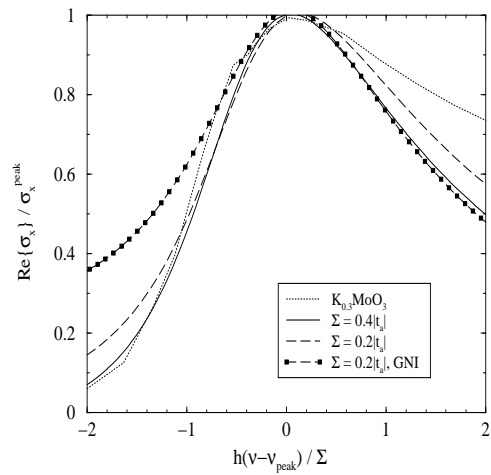}
    \caption{The real part of the optical conductivity
    as a function of the damping energy $\Sigma$, for 
    $|t_b| \rightarrow 0$, $\Delta_0 = 0.5 |t_a|$, and
    $k_{\rm F} = 0.5\pi/a$.
	The filled squares illustrate the prediction 
    of the usual, gauge-non-invariant (GNI) transverse model.
	The experimental data measured in 
    K$_{0.3}$MoO$_3$ (Ref. \cite{Degiorgi2}) 
    are represented by the dotted line.
    }
     \end{figure}

	Fig. 1 shows the influence of the 
dimerization on the (normalized) optical conductivity in the most 
interesting metallic case ($k_{\rm F} = 0.475 \pi/a $),
where the $T \approx 0$ K total spectral weight is shared between the intraband
and interband contributions nearly in equal proportions.
	The participation  of the intraband (i.e. Drude)
part in the total spectral weight 
raises significantly with increasing temperature.
	On the other hand, 
the temperature dependence of the dc  conductivity 
$\sigma_x^{\rm dc} \equiv {\rm Re } \{ \sigma_x^{\rm total} (0) \}$
results from two opposite effects,
from this explicit temperature trend and from the implicit
temperature dependence in $\Sigma (T)$ and/or $\Delta_0 (T)$.	
	The experimental data illustrating this competition,
measured in various low-dimensional systems,
can be found in Refs. \cite{Jacobsen2,Schwartz,Lupi},
while a similar theoretical analysis is given in Ref. \cite{Emery}.

	More importantly, this figure illustrates that the interband 
contribution vanishes for $\omega \rightarrow 0$,
independently of $k_{\rm F}$.
	Such behaviour  is due to the 
gauge-invariance factor $[\hbar \omega / E_{AS} (k_x)]^2$ in 
Eq. (\ref{eq13}).
	The physical meaning of this observation is that 
in the metal-to-insulator phase transitions
(for example, for $k_F \rightarrow 0.5 \pi /a$, $\Delta_0 \neq 0$) 
the dc conductivity  has to vanish.
	The usual transverse approach, 
where the gauge-invariance factor is absent,
gives, on the other hand, a finite $\sigma_x^{\rm dc}$ 
not only in the metallic regime, but also 
in the insulating regime
(see Fig. 2 and an extensive discussion done in Ref. \cite{Kim}). 
	These conclusions are valid not only 
 for the bond-energy-dimerization optical models, 
but also for  the site-energy-dimerization optical models 
\cite{Kupcicup}, 
as well as for the single-particle contributions 
to $\sigma_x^{\rm total} $ in the systems with  CDW or SDW 
instabilities.

To compare briefly the predictions of the present model with experiments, 
we apply it now to the insulating $k_{\rm F} = 0.5 \pi /a$ case, and,
again, normalize the spectra.
	For the parameter $\Delta_0 = 0.5 |t_a|$ describing the 
weak-splitting case, we obtain the results, shown in 
Fig. 2,  which fit well the single-particle optical conductivity 
measured in the CDW ground state of K$_{0.3}$MoO$_3$.
	Notice that the usual transverse model (filled squares) significantly
overestimates the subgap spectrum.

Finally,  it should be recalled that 
the gauge invariant form of $\sigma_x^{\rm total} $
allows the consistent treatment of the conductivity sum rules 
\cite{Kupciccm}.
	For the lower band partially filled,
the spectral weights of the intraband and total conductivity
are given by
 \begin{eqnarray}
\frac{1}{2} \Omega_{\rm intra}^2 &=& \frac{2 \pi e^2 n^{\rm eff}_x}{m},
\nonumber \\
\frac{1}{2} \Omega_{\rm total}^2&=& \frac{2 \pi e^2}{m}
\frac{1}{V} \sum_{{\bf k} \sigma} 
\gamma^{SS}_{xx} ( {\bf k};2)   f_S({\bf k}),
\label{eq19}\end{eqnarray}
respectively,
and that of the interband contributions by
$(1/2)(\Omega_{\rm total}^2 - \Omega_{\rm intra}^2)$.
	Moreover, the dc conductivity is directly related to the intraband 
spectral weight, according to the well-known relation
$\sigma_x^{\rm dc} =  \hbar \Omega_{\rm intra}^2 / (4 \pi \Sigma$).

\subsection{$|t_b| \rightarrow 0$, $T \rightarrow 0$ limit}
The characterization of the low-frequency metallic response is
intimately connected with the energy region between 
$\Sigma_1$ and $ E_{AS} ( k_{\rm F})  - \Sigma_2$
(see Fig. 1, for example).  
	In general, in the weak-splitting optical models, both
the Drude-like behaviour (characterized by
${\rm Re} \{ \sigma_x ( \omega) \} \propto \omega ^{-2}$)
and the non-Drude behaviour 
(where
${\rm Re} \{ \sigma_x ( \omega) \} \propto \omega ^{-\alpha},
\alpha < 2$) are possible,
as can be seen from Fig. 3, where the real part of the 
conductivity for various Fermi wave
vectors is given.

    \begin{figure}[tb]
     \includegraphics[height=15pc,width=15pc]{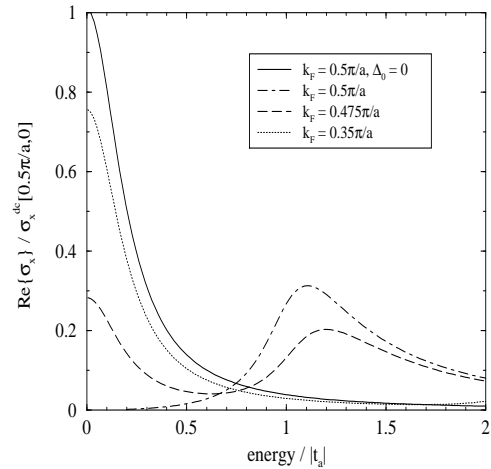}
    \caption{The real part of the optical conductivity    
    as a function of the Fermi wave vector, for
    $\Delta_0 = 0.5 |t_a|$ and  
    $\Sigma  = 0.2 |t_a|$.
	Here $\sigma_{\rm dc} (0.5\pi/a, 0)$ is the 
    dc conductivity of the $k_{\rm F} = 0.5\pi/a$, $\Delta_0 = 0$ case.
    }
     \end{figure}

    \begin{figure}[tb]
     \includegraphics[height=15pc,width=15pc]{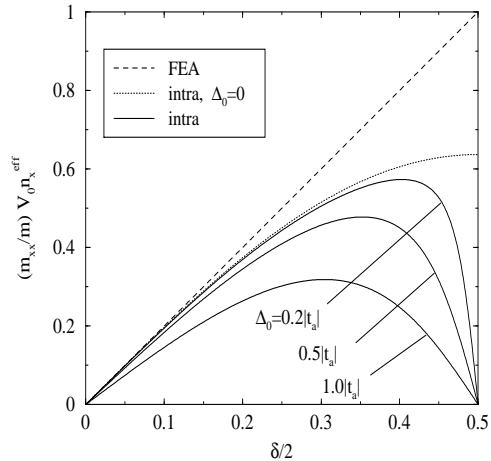}
    \caption{The effective number of conduction electrons
    $V_0n_{x}^{\rm eff}$ as a function of the electron doping 
    $\delta = 2ak_{\rm F}/\pi$ 
    ($V_0$ is the primitive cell volume).
	The prediction of the free-electron approximation
    (FEA) is also given for comparison.	
    }
     \end{figure}

A more detailed insight into the low-frequency response 
can be achieved from the doping dependence of $n_{x}^{\rm eff}$,
which is shown in Fig. 4.
	The following conclusions are important:
	(i) The free-electron approximation (dashed curve) 
predicts the electronlike 
behaviour of the conduction electrons 
($\partial n^{\rm eff}_{x} /\partial \delta  > 0$) in the entire doping range
$0 \le \delta \le 2$.
	(ii) The simple tight-binding model (dotted curve) 
leads to the electronlike regime for $0 \le \delta < 1$ and to the holelike
regime ($\partial n^{\rm eff}_{x} /\partial \delta  < 0$) for
$1 <  \delta \le 2$.
	(iii) The dimerized tight-binding model (solid curves)
is characterized by the electronlike behaviour for 
$0 \le \delta < 1 - \delta_{\rm c}$ and $1  < \delta < 1 + \delta_{\rm c}$,
and by the holelike behaviour  for 
$1 - \delta_{\rm c} < \delta < 1 $ and $1 + \delta_{\rm c} < \delta \le 2$.
	(iv) The non-Drude response is expected in the 
$1 - \delta_{\rm c} < \delta < 1 + \delta_{\rm c}$
doping region, i.e., 
where the spectral weight of the interband contributions
becomes comparable to the intraband spectral weight.

	For $\Delta_0$ not to large, the critical doping $\delta_{\rm c}$
is proportional to $\Delta_0$.
	Not surprisingly, for $1 - \delta_{\rm c} < \delta=1 - \delta' < 1 $ 
the electronic system would be described by the effective mass approximation
which is given by
\begin{eqnarray}
V_0n_{x}^{\rm eff} &=& \frac{m}{m_{xx}^*}\delta'
= -\frac{m}{m_{xx}^*}\delta + \frac{m}{m_{xx}^*}
\label{eq20}
\end{eqnarray}
with $\delta'$ and $m_{xx}^*$ representing, respectively, 
the number of doped holes (with respect to the half filling)
and the effective mass of these holes.
	It must be noted that the effective mass $m_{xx}^*$ is proportional now
to $\Delta_0$ and that in the limit $\Delta_0 \rightarrow 0$ 
this mass  vanishes.

	\begin{figure}[tb]
	 \includegraphics[height=15pc,width=15pc]{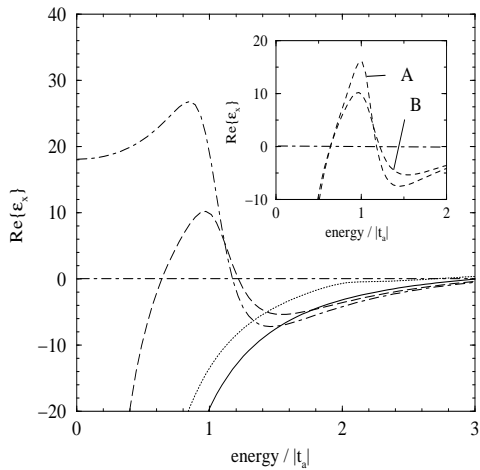}
	\caption{
	Main figure: The dependence of the real part of the dielectric function 
	on the Fermi wave vector,
	for $\Delta_0 = 0.5|t_a|$,
	$\hbar \sqrt{4 \pi e^2 /(V_0m)} = 6 |t_a|$, 
	$\varepsilon_{\infty} = 2.5$, 
	and $\Sigma = 0.2 |t_a|$
	(see the legend of Fig 3).
	Inset of figure: The influence of the damping energies
	on ${\rm Re}\{ \varepsilon_x(\omega )\}$, 
	for $k_{\rm F} = 0.475 \pi/a$, $\Sigma = 0.12 |t_a|$ (curve A),
	and $\Sigma = 0.2 |t_a|$ (curve B).
		}
	 \end{figure}

Finally, the real part of the dielectric function is 
affected by the dimerization in the way shown in Fig. 5.
	In the Drude regime the influence of the
dimerization is negligible, giving 
only one zero-crossing of
${\rm Re}\{ \varepsilon_x(\omega )\}$ placed at the frequency 
of the longitudinal intrachain ($q_y = 0$ in Eq. (\ref{eq18})) plasma mode. 
	In the non-Drude regime 
two zeros of the dielectric function appear, describing  
the intrachain plasma oscillations, as well.
	The first frequency 
corresponds to the intraband  plasma oscillations
and the latter one to the total plasma oscillations.
	As illustrated in the inset of figure,
in the underdamped regime ($\Sigma_i < \Delta_0$),
considered in this article,
the zeros of ${\rm Re}\{ \varepsilon_x(\omega )\}$ 
are almost independent of the single-particle damping.

\section{Conclusion}
In this paper we have studied the influence of the weak bond-energy 
dimerization on the optical properties of the Q1D systems, for the 
electron doping $0 \le \delta \le 1$.
	It is found that the gap parameter has the $p_x$ symmetry.
	It is shown that  the weak bond-energy dimerization 
leads to  the dielectric function in which the 
local field corrections are negligible.
	The analysis reveals two different metallic states. 
	First state corresponds to the usual electronlike Drude metal
with the optical conductivity 
${\rm Re}\{ \sigma_x ( \omega) \} \propto \omega^{-2}$,
with the effective number of electrons characterized by
$\partial n^{\rm eff}_{ x} /\partial \delta > 0$ and only with one zero-point
in ${\rm Re}\{ \varepsilon_x ( \omega) \} $.
	Second metallic state is the non-Drude state characterized
by the $\omega^{-\alpha}$ behaviour  of the low-frequency
contribution to  ${\rm Re}\{ \sigma_x ( \omega) \}$, where $\alpha < 2$.
	There are two zero-points in ${\rm Re}\{ \varepsilon_x ( \omega) \} $,
and the effective number of electrons exhibits the holelike behaviour
$\partial n^{\rm eff}_{ x} /\partial \delta < 0$.
	The comparison with experimental data measured in various 
CDW systems shows good agreement for the single-particle 
contributions to the optical conductivity.

\section*{Acknowledgement}
This work was supported by Croatian Ministry of Science under the 
project 119-204.

\appendix
\section{Vertex functions}

	For the lower band  partially filled and the upper band empty, 
 relevant are the current vertices (again $r_{\beta x} = a$)
\begin{eqnarray}
J^{SS}_{x} ( k_x) &=&  J_x^0
[ \sin  k_x  a \cos \varphi ( k_x) 
\nonumber \\ 
&&-
\frac{\Delta_0}{2|t_a|} \cos  k_x a \sin \varphi ( k_x)] ,
\nonumber \\
J^{SS}_{y} ( k_x) &=&  J_y^0 \sin  k_y  b,
\label{eq21} \\
J^{SA}_{x} ( k_y) &=&   \mathrm{i} J_x^0
[\sin  k_x a \sin \varphi (k_x) 
  \nonumber \\ \nonumber 
&&+ \frac{\Delta_0}{2|t_a|}
\cos  k_x a \cos \varphi (k_x)],
\nonumber \\
J^{SA}_{y} ( k_y) &=& 0, 
\label{eq22}\end{eqnarray}
and the bare Raman vertices 
\begin{eqnarray}
\gamma ^{SS}_{xx} ( k_x;2) &=& 
\gamma_{xx}^0 
[  \cos  k_x a \cos \varphi ( k_x) 
\label{eq23} \\ \nonumber
 &&+
 \frac{\Delta_0}{2|t_a|} 
\sin k_x a \sin \varphi ( k_x)],
\nonumber \\
\gamma ^{SS}_{yy} (k_y ;2) &=& \gamma_{yy}^0 
\cos  k_y b.
\label{eq24}\end{eqnarray}
	Here $J_{x_i}^0 = e \hbar/(a_i m_{x_i x_i})$, 
$\gamma_{x_i x_i}^0 = m/m_{x_i x_i}$ and 
$m_{x_i x_i} = \hbar^2/(2|t_i|a_{i}^2)$
($|t_i| \in \{ |t_a|,  |t_b| \}$, 
$a_i \in \{ a, b \}$).

\end{document}